# Segmentation of Pediatric Brain Tumors using a Radiologically informed, Deep Learning Cascade

Timothy Mulvany[1*], Daniel Griffiths-King[1*][0000-0001-5797-9203], Jan Novak[1◊][0000-0001-5173-3608] & Heather Rose[2◊][0000-0002-0346-1334]

[1]Aston Institute of Health and Neurodevelopment, Aston University, Birmingham, UK
[2]College of Engineering and Physical Sci., Aston University, Birmingham, UK
`j.novak@aston.ac.uk`

*These authors contributed equally as joint first authors.
◊These authors contributed equally as joint senior authors.

**Abstract.** Monitoring of Diffuse Intrinsic Pontine Glioma (DIPG) and Diffuse Midline Glioma (DMG) brain tumors in pediatric patients is key for assessment of treatment response. Response Assessment in Pediatric Neuro-Oncology (RAPNO) guidelines recommend the volumetric measurement of these tumors using MRI. Segmentation challenges, such as the Brain Tumor Segmentation (BraTS) Challenge, promote development of automated approaches which are replicable, generalizable and accurate, to aid in these tasks. The current study presents a novel adaptation of existing nnU-Net approaches for pediatric brain tumor segmentation, submitted to the BraTS-PEDs 2024 challenge. We apply an adapted nnU-Net with hierarchical cascades to the segmentation task of the BraTS-PEDs 2024 challenge. The residual encoder variant of nnU-Net, used as our baseline model, already provides high quality segmentations. We incorporate multiple changes to the implementation of nnU-Net and devise a novel two-stage cascaded nnU-Net to segment the substructures of brain tumors from coarse to fine. Using outputs from the nnU-Net Residual Encoder (trained to segment CC, ED, ET and NET tumor labels from T1w, T1w-CE, T2w and T2-FLAIR MRI), these are passed to two additional models one classifying ET versus NET and a second classifying CC vs ED using cascade learning. We use radiological guidelines to steer which multi parametric MRI (mpMRI) to use in these cascading models. Compared to a default nnU-Net and an ensembled nnU-net as baseline approaches, our novel method provides robust segmentations for the BraTS-PEDs 2024 challenge, achieving mean Dice scores of 0.657, 0.904, 0.703, and 0.967, and HD95 of 76.2, 10.1, 111.0, and 12.3 for the ET, NET, CC and ED, respectively.

# 1   Introduction

## 1.1   Background

Monitoring of Diffuse Intrinsic Pontine Glioma (DIPG) and Diffuse Midline Glioma (DMG) brain tumors in pediatric patients using magnetic resonance imaging (MRI) is key for diagnosis, disease prognosis, and assessment of treatment response. For instance, Response Assessment in Pediatric Neuro-Oncology (RAPNO) guidelines recommend the volumetric measurement of these tumors using MRI. Segmentation challenges, such as the Brain Tumor Segmentation (BraTS) challenges[1], promote development of automated approaches which are replicable, generalizable, and accurate, to aid in these tasks.

In the BraTS 2022 Glioma challenge, submitted segmentation models trained on adult data, were applied to a held-out cohort of pediatric data [1]. The poor performance achieved by these models when applied to a child cohort of pediatric brain tumors emphasized the need for deep learning and machine learning models that are specifically trained on pediatric data.

The current study presents a novel adaptation of existing nnU-Net approaches for pediatric brain tumor segmentation, submitted to the BraTS-PEDs 2024 challenge. We apply an adapted nnU-Net with hierarchical cascades to the segmentation task of the BraTS-PEDs 2024 challenge.

## 1.2   Previous Work

**nnU-Net**. nnU-Net is an adaptive, deep-learning segmentation approach [2]. It is adaptive in that configuration of the model (for instance preprocessing, network architecture, training, and post-processing), is automatically updated in response to the training data supplied. This out of the box and adaptable approach, provide an impressive baseline which, through additional experimentation, is a strong starting point for challenge-specific optimization and extension [3].

nnU-Net featured heavily in the BraTS-PEDs 2023 challenge, of the 9 reported entries, 3 used nnU-Net (or a derivative thereof) to segment pediatric brain tumors, with two instances appearing in the top 3 performing models [1]. The second ranking team used an adaptation of nnU-Net, with self-supervised pretraining integrated with adaptive region-specific loss [1]. A team containing challenge organizers (which therefore went unranked) also used nnU-Net, ensembled with Swin UNETR. Whilst unranked, the model performed similarly to the winning entry, with no significant difference in performance [1].

An adapted version of nnU-net using asymmetrically adapted network size, batch normalization and an attention decoder achieved first rank in the 2021 BraTS Challenge for segmenting adult glioma [3]. A task optimized nnU-Net also achieved first place at the BraTS 2020 challenge [4] with adaptations to the nnU-net default configuration including increased batch size, batch (versus instance) normalization, batch (versus sample) Dice loss.

---

[1] https://www.synapse.org/Synapse:syn53708249/wiki/626323

The current study utilizes nnU-Net as its underlying deep-learning architecture for model design, proposing an adaptation for pediatric brain tumor segmentation, whilst additionally using nnU-Net Residual Encoder for experimentation as an effective baseline model. This is a suitable baseline given recent guidelines proposed by nnU-Net [5] and ablation studies demonstrating the clear benefit of residual blocks on performance on brain tumor segmentation tasks in both adults and pediatrics [5].

**Cascade Models** Recent narrative review highlights the number of cascade learning approaches for the task of tumor segmentation, across solid tumors from multiple organ systems (e.g. liver, kidney etc.) [6]. Cascade learning involves training a base model and using the 'knowledge encapsulated in that prediction, alongside the original input images, in a subsequent model [7] (in the current study we refer to base and subsequent models as Stage 1 and Stage 2 models respectively).

In ablation experiments of their model Chen and colleagues [8], demonstrated that a 2-stage cascaded 3D U-Net achieves finer segmentation results than a baseline 3D U-Net, across both BraTS 2020 training and validation cohorts. In brain tumor segmentation, cascade learning can also take advantage of the hierarchical nature of labels used to describe tumor subregions. Wang et al [9] used a cascade of CNNs for sequential segmentation of brain tumor and subregions from multimodal MRI. They used a bounding box and crisp masking approach to combine inputs from previous levels of the cascade into next stages. However, this can allow the propagation of error in the initial first stage segmentation to impact later segmentations further down the cascade. In breast tumor segmentation, a 2 stage cascaded model performed course to fine segmentation in a hierarchical fashion, with improved segmentation which perform well in terms of a downstream classification task [10].

A previous study utilized a cascade architecture to improve performance of their CNN which used a 2 phase training procedure for learning label class imbalance and both local and global convolutional pathways for low- and high-grade glioma segmentation from MRI [10]. They tested at which point concatenation of the output of the first CNN with the second should occur; finding that inputting these outputs directly as additional input modalities to the second model outperformed both a) concatenating with the first hidden layer of the second model, or b) concatenating before the first output layer of the second model. They outperformed winning entries on the BraTS 2013 and 2012 challenges and performed competitively in the BraTS 2015 challenge.

**Current Study** Studies using cascade learning such as Wang et al [9], are also able to benefit from the decomposition of the complex segmentation task presented by brain tumors, into simpler, binary tasks (such as Whole Tumor (WT) vs background, then Tumor Core (TC) vs Background). Given the additional sub-regions required for the BraTS-PEDs 2024 challenge (Enhancing Tumor (ET), Non-enhancing Tumor (NET), Cystic Component (CC) and Edema (ED)) binary classification would require a greater number of classifiers to be trained, being resource intensive. This approach is required if the work is to be translated into routine clinical practice.

Instead, we propose a hierarchical set of cascaded classifiers, focusing on the boundaries for which the classifications are more difficult, both in automatic segmentation, but also informed by difficulties raised in manual radiological segmentation. After an initial classification model (Stage 1 Model) classifying all tumor subregions, we designed two subsequent models, one classifying ET versus NET and a second classifying CC vs ED. Through integration with radiological guidelines, specifically RAPNO guidelines for DIPG and HGG [11-13] we are able to make radiologically-guided decisions on appropriate input MRI modalities for these cascaded models. This enables us to decompose the initial segmentation into two simpler models, with fewer input modalities and classification labels - whilst not being resource intensive like the four binary classifiers which would be needed to capture all 4 subregions.

## 2 Methods

### 2.1 Data

Data used in this submission were obtained as part of the Challenge project through Synapse ID (syn51156910).
**Subjects.** The BraTS-PEDs dataset consists of data from n=464 patients with pediatric high-grade glioma (e.g. high-grade astrocytoma, DMG and DIPG) from multiple institutions. n=261 cases were used in the training phase, with MRI and data labels being available. An additional n=91 cases were used for validation testing of the models, with only MRI being made available. Final benchmarking of our submission will take place on the final testing cohort of n=112, where neither MRI nor labels have previously been released. See [14] for further details
**MRI.** The BraTS-PEDs dataset contains a multi-institutional, retrospective cohort of multiparametric MRI (mpMRI) sequences. For each patient, multiple MRI modalities were available: native (T1w), postcontrast T1-weighted (T1w-CE), T2-weighted (T2w), and T2 Fluid Attenuated Inversion Recovery (T2-FLAIR) modalities.
**Data Annotations – Tumor Sub-regions.** Annotated "truth" tumor masks were used for model training and included labels for four tumor sub-regions: "enhancing tumor" (ET), "non-enhancing tumor" (NET), "cystic component" (CC) and "edema" (ED). These could be meaningfully combined to generate three additional labels, "tumor core" (TC) combining ET, NET, and CC, and finally the "whole tumor" (WT) - the entire tumorous region combining ET, NET, CC and ED.
**MRI Preprocessing** MRI Data was preprocessed using a standardized publicly-available approach, termed the "BraTS Pipeline", through the Cancer Imaging Phenomics Toolkit (CaPTk) and Federated Tumor Segmentation (FeTS) tool, and anonymized through removing protected DICOM headers and defacing of MRI (see [14] for details). Additional preprocessing was conducted, specifically in terms of patient-level normalization. For each MRI modality, for every case, voxel intensities from within brain tissue only (estimated using a mask generated using a python implementation[2] of FSL BET [15]), were binned forming a histogram to which a Gaussian was fitted to the greatest peak. Voxel intensities were divided by two times the mean parameter of the

---

[2] https://github.com/vanandrew/brainextractor.git

Gaussian curve, to normalize the Gaussian peak to 0.5. This was to ensure normalization at the individual-level as nnU-Net only performed normalization at a cohort/dataset level.

## 2.2  Model Architectures

**nnU-Net.** In this work we implement multiple configurations of nnU-Net [4]. nnU-Net is an adaptive, deep-learning segmentation approach. It is adaptive in the sense that the configuration (for instance preprocessing, network architecture, training, and post-processing), is automatically updated in response to the training data supplied. Three nnU-Net architectures were designed to develop the current method; i) nnU-Net Residual Encoder, ii) Multi nnU-Net Ensemble and iii) nnU-Net Cascade. For all Model architectures, training was conducted using a 5-fold cross validation approach. For evaluation on validation and testing data, these models were ensembled across the folds, using the default nnU-Net behavior.

1. **nnU-Net Residual Encoder.** As per the guidelines offered by nnU-Net, the nnU-Net residual encoder variant was used as our benchmarking model[3] [5]. Specifically, the "nnU-Net ResEnc M" preset was used, due to its similar GPU requirements to standard UNet configurations. The residual encoder model uses residual blocks in the encoder which perform the function of adding the convolution block's input to the output, preserving information from the previous layers. The inputs to this model are the four mpMRI scans (T1w, T1w-CE, T2-FLAIR and T2w), whilst the output is the four independent tumor subregions (ET/NET/CC/ED).
2. **Multi nnU-Net Ensemble.** To leverage the benefits of multiple configurations of the 3D nnU-Net, we ensembled three nnU-net variants; firstly, a default iteration of nnU-Net, the nnU-Net residual encoder variant described above, and finally a low-resolution variant whereby the image resolution was reduced (by an approximate factor of 1.5) whilst maintaining the physical patch size. The outputs of these three models were then ensembled for inference. This ensembling approach, over multiple variants/configurations of nnU-Net, has performed well on previous MRI and CT segmentation challenges (not BraTS) [16]. The inputs to all models within the ensemble are the four mpMRI scans (T1w, T1w-CE, T2-FLAIR and T2w), whilst the output is the four independent tumor subregions (ET/NET/CC/ED).
3. **nnU-Net Cascade.** The final model architecture used a hierarchical cascade approach, where the outputs of an initial model (Stage 1 models) are passed to subsequent models as an additional input channel (Stage 2 models) for refinement. The Stage 1 model is the nnU-Net Residual Encoder (as the strongest performing individual model – between the default, low resolution, nnU-Net Residual Encoder and Multi nnU-Net Ensemble). The input to Stage 1 is the four mpMRI scans (T1w, T1w-CE, T2-FLAIR and T2w), whilst the output is the four independent tumor subregions (ET/NET/CC/ED). A further two models were trained in a cascade framework, with outputs of the Stage 1 model fed forward to the Stage 2 model. Given these two further models in the cascade are performing similar tasks in terms of their level/position in the cascade hierarchy (in terms of inputs and outputs), we termed them both stage 2a (ET vs NET segmentation) and stage 2b (CC vs ED segmentation). We also reduced input channels by specifically selecting, out of the four available mpMRI sequences, the modalities that were most appropriate to the subregions being segmented by that stage 2 model, selected based upon RAPNO guidelines.

---

[3] As per https://github.com/MIC-DKFZ/nnUNet/blob/96253e9dae2e7879520ab5dfc1c84aefe0f818e7/documentation/resenc_presets.md

The Stage 2a model (ET vs NET segmentation) is generated using the default nnU-Net configuration. The input to this model is the ET and NET label predictions from the Stage 1 model, plus the T1w and T1w-CE MRI. The output labels are ET and NET. For the stage 2b model (CC vs ED segmentation), the ET and NET labels from the Stage 1 model are combined to generate a "Solid Tumor" label (ST). The ST label, plus CC and ED labels from the Stage1 model, are passed as inputs to the Stage 2b model as well as the T2w and T2-FLAIR MRI. The outputs are refined CC and ED labels. Stage 2a model represents a cascade to refine prediction, whereas Stage 2b represents refining segmentations from coarse (ST) to fine (CC/ED).

Label predictions from the Stage 2 models (a & b) are combined with a simple ensembling. No collisions between ET/NET or between ED/CC can exist because of the output space, but the current approach creates a final prediction by having Stage 2b (CC + ED) outputs overwrite Stage 2a (ET + NET) outputs.

### 2.3   Training & Evaluation.

We followed the standard training methodology of nnU-Net for all architectures. Each architecture was trained with 5-fold cross-validation. nnU-Net applied default data augmentation on-the-fly during training. Each Training run lasted 250 epochs with a batch size of 2. The networks were optimized with stochastic gradient descent with Nesterov momentum of 0.99. The initial learning rate was 0.01 and was decayed following a polynomial schedule. All experiments were conducted with Pytorch (v2.3.0 + CU12.1) on two NVIDIA Quadro RTX 6000 GPUs with 24GB VRAM. Combination of predictions from differing models / folds was performed using python 3.9.19, nibabel 5.2.1.

Performance is reported for each model on the BraTS-PEDs 2024 validation set. This dataset is made available to challenge participants without ground truth tumor labels. Performance on validation data for this challenge was assessed/benchmarked using the synapse.org platform, powered by MedPerf. [17].

## 3   Results

Results report performance on the validation dataset unless explicitly stated. Example segmentations of final model can be seen in figure 1.

**Ensembling multiple nnU-Net configurations erodes segmentation performance.** The Multi nnU-Net Ensemble in which we ensembled nnU-Net runs both across folds (in our 5-fold cross-validation schema) and across multiple configurations (Default,

Low resolution and Residual Encoder nnU-Net) did not outperform the Residual Encoder baseline on all labels, other than CC (Table 1). Given that the Ensemble included the Residual Encoder, this suggests that the addition of other models eroded the higher performance of the Residual Encoder alone.

A previous iteration of the cascade used the Multi nnU-Net Ensemble as the Stage 1 model, however, in light of a) the Residual Encoder outperforming the Multi nnU-Net Ensemble and b) the nnU-Net Cascade performing better with the Residual encoder as the Stage 1 model versus the ensemble as Stage 1, our nnU-Net Cascade used the Residual Encoder as the Stage 1 model.

**nnU-Net Cascade provides small benefit in CC and ED delineation.** Both nnU-Net Cascade and nnU-Net Residual Encoder perform similarly on ET and NET (and combined TC and WT masks). The (negligible) difference in average Dice for ET and NET are 0.004 and 0.002 respectively in the direction of the Residual Encoder. However, the mean Dice for CC and ED show a 0.013 and 0.011 slight improvement for the nnU-Net Cascade model. For these reasons the nnU-Net Cascade was chosen to be submitted for evaluation in the challenge.

**Improvements in CC for nnU-Net Cascade due to Increased True Negative Detection.** Further review of the nnU-Net Cascade indicated all validation cases to be predicted to have had no cystic component, whereas the nnU-Net Residual Encoder model had predicted CC in cases where it had been non-existent, with these false positives having a greater effect than the false negatives in the cascade model. When we assessed the performance on the training data, the nnU-Net Residual Encoder model predicted "empty" masks for CC and ED in 111/261 and 136/261 cases respectively, whilst for the nnU-Net Cascade model this was 156/261 and 103/261. For CC the nnU-Net Cascade is performing more closely to the "empty" mask rate of the Truth masks, perhaps explaining the increased performance.

**nnU-Net Cascade and nnU-Net Residual Encoder models show poor sensitivity.** The nnU-Net Cascade model did not predict ANY CC for the validation cohort, whereas the nnU-Net Residual Encoder predicted CC segmentations in 8 cases, which were true positives, however segmentation performance on these was still low (Dice=0.47), and likely does not represent the true positive rate of the validation cohort for CC. This performance indicates low sensitivity – the ability for either model to detect and correctly segment true positive CC (although the nnU-Net Residual Encoder did identify some cases of CC).

Table 1. Performance of each model on BraTS-PEDs 2024 validation dataset

| Model | Lesionwise Dice | | | | | |
|---|---|---|---|---|---|---|
|  | ET | TC | WT | NETC | CC | ED |
| nnU-Net Residual Encoder | 0.661 | 0.934 | 0.934 | 0.906 | 0.690 | 0.956 |
| Multi nnU-Net Ensemble | 0.616 | 0.931 | 0.931 | 0.903 | 0.729 | 0.901 |
| nnU-Net Cascade | 0.657 | 0.931 | 0.931 | 0.904 | 0.703 | 0.967 |

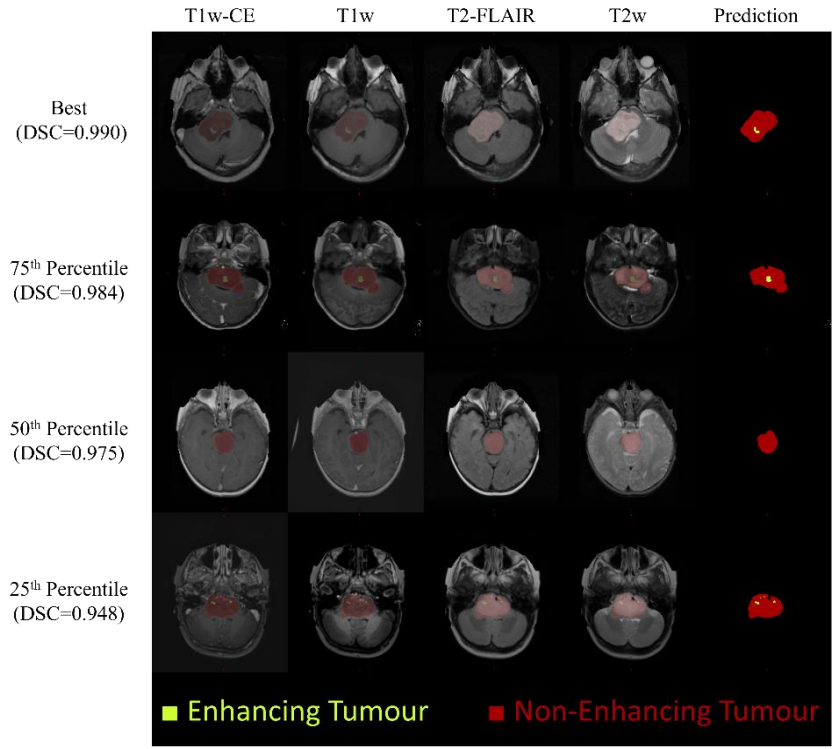

**Fig. 1.** Example predictions of decreasing quality from the validation cohort using the nnU-Net Cascade (based on average TC and WT Dice). ET is overlayed in yellow, and NET in red. Note predictions from these cases did not include CC or ED in our model.

## 4    Discussion

The current paper outlines a novel adaptation of the nnU-Net segmenter for the purposes of pediatric brain tumor segmentation. We propose a two-stage cascaded nnU-Net, as a submitted solution to the ASNR-MICCAI BraTS Pediatrics Tumor Challenge 2024 [14]. Our approach refines the prediction through a progressive cascaded network, which is radiologically informed by both the hierarchical information between tumor subregions, but also in the specific mpMRI modalities provided at each stage of the cascade. We note that the approach only demonstrates improvement, compared to the nnU-Net residual encoder baseline, on CC and ED labels. However, difficulties in cyst segmentation are a common issue identified by challenge organizers [14],

Whilst both the residual encoder and cascade performed similarly across most of the tumor subregions, the cascade performed better on the cystic component. Through further investigation, we believe this to be due to increased detection of true negative cases – the cascade was better at identifying when there was no cyst in the image (thus predicting an empty CC label) than the residual encoder model, which predicted more false-positive CC.

However, in the validation cohort, it is noted that this resulted in all cases being predicted as having no cystic component. In the absence of cysts, the model performs well in the challenge, achieving a Dice score of 1. For cases where there is cyst which has been missed in the segmentation, a Dice score of 0 is assigned. This behavior, if repeated in the test set, may indicate that imbalances in the training set provided challenges for training our model [18]. Class imbalances existed in the CC labels within scans (i.e. greater proportion of background plus other labels to CC) but also across the cohort, in the training data (and potentially in the validation set). Our models have focused on the segmentation of features in the scan and therefore is not designed specifically to triage for feature presence. This should be addressed for future models given that in real-world data sets are also likely to be imbalanced [18].

Whilst this does not necessarily present an issue for the current challenge, under- or non- segmentation of cyst and edema has clinical implications – both are important radiological features in pediatric brain tumors [18, 19] and RAPNO guidelines highlight excluding some cysts in response assessment [12]. Therefore, the current approach, despite being radiologically informed, may not be best suited to actual clinical challenges. In unreported experiments we tried to improve sensitivity by stratify training of our cascade models based upon presence of CC and ED, training only on cases that had cystic or edema within the image. However, this still performed poorly with a much higher false positive rate in the validation data (trying to segment CC when it did not exist).

Our results also showed that our model did not improve segmentation of NET and ET compared to the Residual Encoder baseline. This could imply that it is possible to simplify our network by removing the initial Stage 2a pathway. However, we would argue that dual-pathway cascade utilized in the current project has not fully been exploited due to time constraints of the challenge itself. Through inclusion of this branching design of the cascade, as well as simplifying the segmentation task, further radiological domain knowledge can be incorporated at individual stages using additional image processing techniques, as has been previously demonstrated resulting in improved segmentation [20]. Future work will try to leverage this dual-pathway or branching cascade design to this effect.

Another potential difficulty in segmentation raised by challenge organizers, was the segmentation of NET as CC (and vice versa) [14]. In designing our models, we attempted to address this in two ways. Firstly, our cascade was designed in a way to isolate final CC and NET predictions into two different stage 2 classifiers (Stage 2a and 2b), preventing direct collisions when ambiguity is high between the two classes. Also, in ensembling the outputs of the two stage 2 models, we overwrote any collisions between the two outputs with the output of the Stage 2b (ET + NET) model (chosen because the Stage 2a model performed slightly more accurately in experimentation and validation). This allows us to potentially overwrite misclassifications where True NET is segmented as CC (when Stage 2a has correctly identified the True NET). However, this does not necessarily improve where True CC is incorrectly labelled as NET, or if the Stage 2a model does not correctly identify True NET. This may explain our poorer performance in CC. Future experimentation should also assess whether a further stage 2 (or even 3) model, explicitly trained to discriminate NET from CC, to further refine predictions.

nnU-Net stands for "no new U-Net" [2]. The authors, and related scholars, propose that new network architectures, or even expansion of existing approaches, are not required to achieve state of the art segmentation performance. In fact, their results suggest that efficient and accurate optimization, automated in the nnU-Net framework, provides comparable or higher performance than many specialist pipelines. Hence their recommendations that the residual encoder variant is a strong and effective baseline for future experimentation [5]. It may therefore be unsurprising that, as a single adaptation of the model, our expansion of the nnU-Net, using radiologically informed, hierarchical classification, did not provide great improvements in performance. However, our contributions are two-fold, a) on current validation data, our model has improved CC and ED and b) our radiologically informed, dual pathway cascade design could be further refined to identify other radiological rules for future performance increases.

**Strengths and Limitations.** A relative strength of the current evaluation is that due to the automated optimization of the nnU-Net configuration, this provides a "strong" baseline with which to compare our proposed model. Thus, our performance relative to baseline is not overstated, due to an underperforming, poorly optimized baseline model.

Limitation of some cascaded models for tumor segmentation is using segmentations at earlier stages of the hierarchy, to mask either a) input modalities for subsequent segmentation or b) post processing of predictions. For instance, Wang et al [9] used their WT segmentations from the first stage of their cascaded model as a bounding box to mask multimodal MRI inputs for the $2^{nd}$ stage model, segmenting TC. They also masked their ET predictions using TC masks. Both of these approaches place strong spatial constraints on segmentations – guided by the hierarchical-like anatomical constraints of tumor subregions. However, in the current challenge, with segmentations of WT, TC, NET, TC, CC and ED, these hierarchical constraints are more complex and do not necessarily hold. Additionally, inclusion of base classifiers errors, in with original data, can amplify and propagate these errors in subsequent stages and predictions. Hence, we passed predictions from higher in the cascade to later learners as an additional input modality, ensuring that there are no strong-anatomical constraints on predictions and prevent early segmentation errors being propagated throughout the model.

**Acknowledgments.** TM is funded by a PhD studentship from the "Help Harry Help Others" charity and Aston Institute of Health and Neurodevelopment, Aston University. DGK is funded by a post-doctoral fellowship awarded to DGK and JN from the College of Health and Life Sciences, Aston University. HR is funded by Innovate UK and the West Midlands Combined Authority through the West Midlands Healthtech Innovation Accelerator.

**Disclosure of Interests.** The authors have no competing interests to declare that are relevant to the content of this article.

## 5    References

1. Kazerooni, A.F., et al., *BraTS-PEDs: Results of the Multi-Consortium International Pediatric Brain Tumor Segmentation Challenge 2023.* arXiv preprint arXiv:2407.08855, 2024.
2. Isensee, F., et al., *nnU-Net: a self-configuring method for deep learning-based biomedical image segmentation.* Nat Methods, 2021. **18**(2): p. 203-211.
3. Luu, H.M. and S.-H. Park. *Extending nn-UNet for Brain Tumor Segmentation*. in *Brainlesion: Glioma, Multiple Sclerosis, Stroke and Traumatic Brain Injuries*. 2022. Cham: Springer International Publishing.
4. Isensee, F., et al. *nnU-Net for Brain Tumor Segmentation*. in *Brainlesion: Glioma, Multiple Sclerosis, Stroke and Traumatic Brain Injuries*. 2021. Cham: Springer International Publishing.
5. Isensee, F., et al., *nnu-net revisited: A call for rigorous validation in 3d medical image segmentation.* arXiv preprint arXiv:2404.09556, 2024.
6. Jiang, H., Z. Diao, and Y.-D. Yao, *Deep learning techniques for tumor segmentation: a review.* The Journal of Supercomputing, 2021. **78**(2): p. 1807-1851.
7. de Zarzà, I., et al., *Cascading and Ensemble Techniques in Deep Learning.* Electronics, 2023. **12**(15).
8. Chen, B., et al., *Adaptive cascaded transformer U-Net for MRI brain tumor segmentation.* Phys Med Biol, 2024. **69**(11).
9. Wang, G., et al., *Automatic Brain Tumor Segmentation Based on Cascaded Convolutional Neural Networks With Uncertainty Estimation.* Front Comput Neurosci, 2019. **13**: p. 56.
10. Zhang, J., et al., *Hierarchical Convolutional Neural Networks for Segmentation of Breast Tumors in MRI With Application to Radiogenomics.* IEEE Trans Med Imaging, 2019. **38**(2): p. 435-447.
11. Erker, C., et al., *Response assessment in paediatric high-grade glioma: recommendations from the Response Assessment in Pediatric Neuro-Oncology (RAPNO) working group.* Lancet Oncol, 2020. **21**(6): p. e317-e329.
12. Cooney, T.M., et al., *Response assessment in diffuse intrinsic pontine glioma: recommendations from the Response Assessment in Pediatric Neuro-Oncology (RAPNO) working group.* Lancet Oncol, 2020. **21**(6): p. e330-e336.
13. Bhatia, A., et al., *Review of imaging recommendations from Response Assessment in Pediatric Neuro-Oncology (RAPNO).* Pediatr Radiol, 2023. **53**(13): p. 2723-2741.
14. Kazerooni, A.F., et al., *The Brain Tumor Segmentation in Pediatrics (BraTS-PEDs) Challenge: Focus on Pediatrics (CBTN-CONNECT-DIPGR-ASNR-MICCAI BraTS-PEDs).* arXiv preprint arXiv:2404.15009, 2024.
15. Jenkinson, M., et al., *Fsl.* Neuroimage, 2012. **62**(2): p. 782-790.
16. Isensee, F., et al. *Extending nnu-net is all you need*. Springer.
17. Karargyris, A., et al., *Federated benchmarking of medical artificial intelligence with MedPerf.* Nat Mach Intell, 2023. **5**(7): p. 799-810.

18. Familiar, A.M., et al., *Towards Consistency in Pediatric Brain Tumor Measurements: Challenges, Solutions, and the Role of AI-Based Segmentation.* Neuro Oncol, 2024.
19. Resende, L.L. and C. Alves, *Imaging of brain tumors in children: the basics- a narrative review.* Transl Pediatr, 2021. **10**(4): p. 1138-1168.
20. Kotowski, K., et al. *Infusing Domain Knowledge into nnU-Nets for Segmenting Brain Tumors in MRI.* Springer.